\documentstyle[12pt]{article}
\begin{document}
\newcommand{\bea}{\begin{eqnarray}}
\newcommand{\eea}{\end{eqnarray}}
\newcommand{\be}{\begin{equation}}
\newcommand{\ee}{\end{equation}}
\newcommand{\non}{\nonumber}
\global\parskip 6pt
\begin{titlepage}
\begin{center}
{\Large\bf Combinatorial Invariants from Four }\\
\vskip .25in
{\Large\bf Dimensional Lattice Models: II}\\
\vskip .5in
Danny Birmingham \footnote{Supported by Stichting voor Fundamenteel
Onderzoek der Materie (FOM)\\
Email: Dannyb@phys.uva.nl}     \\
\vskip .10in
{\em Universiteit van Amsterdam, Instituut voor Theoretische Fysica,\\
1018 XE Amsterdam, The Netherlands} \\
 \vskip .50in

Mark Rakowski\footnote{Email: Rakowski@yalph2.bitnet}   \\
\vskip .10in
{\em Yale University, Center for Theoretical Physics,\\ New Haven,
CT 06511, USA}  \\
\end{center}
\vskip .10in
\begin{abstract}
We continue our analysis of the subdivision properties of certain
lattice gauge theories based on the groups $Z_{2}$ and $Z_{3}$, in
four dimensions. We prove that the partition function for a closed
four dimensional manifold is unity, at the special subdivision
invariant points. For the case of manifolds with boundary, we show
that Alexander type $2$ and $3$ subdivision of a bounding simplex
is equivalent to the insertion of an operator which is equal to a
delta function on trivial bounding holonomies.
\end{abstract}
\vskip .5in
\begin{center}
YCTP-P11-93 \\
May 1993
\end{center}
\end{titlepage}

\section{Introduction}
In \cite{top1,top2}, a class of lattice gauge theories was introduced
which enjoyed certain topological features. An analytic
proof of the invariance of the Boltzmann weights under all type $(k,l)$
subdivision moves was presented in \cite{top2}.
The fundamental identity established there
specified that a certain product of six Boltzmann weights was unity.
Triviality of the invariant for closed $4$-manifolds, which
we establish in this paper,
is based upon the observation that one can always realize
a closed four dimensional complex as the base of a five dimensional cone.
The Boltzmann weight for the boundary of such a cone is precisely 1 by
this identity, and the result immediately follows. While the theory
essentially reduces to behavior on the 3-boundary, the
novelty of the construction is that gauge invariance actually requires
a 4-dimensional perspective.

With this knowledge in hand, we turn our attention to the case
of four manifolds with boundary. The central problem is to determine
how the partition function behaves with respect to subdivision of
the boundary, and we analyze the theory under the Alexander moves
of type $2$
and $3$. We show that subdivisions of these types are equivalent to the
insertion of an operator in the partition function which is a delta
function on trivial bounding holonomies. While it is intriguing that
subdivision of the underlying lattice naturally induces these constraints,
the construction of an object which is fully subdivision invariant will
not be resolved here.

In the following section, we briefly recall the definition of the
Boltzmann weight, together with the statement of the fundamental
identity. The triviality for closed four manifolds is then
established. Considerations of boundary subdivision follow, and we present
our analysis of the Alexander moves of type $2$ and $3$.
We close with a few remarks.

\section{Cobordism Independence}

   We show in this section that the four dimensional models that we have
been considering actually reduce to behavior on the boundary.
In particular,
the partition function is identically 1 for all closed 4-manifolds.

Let us first recall that these models are defined by a Boltzmann
weight $W$, which is a product of $15$ distinct factors $B$; the
generic structure of $B$ is given by:
\bea
B[v_{0},v_{1},v_{2},v_{3},v_{4}] =
\exp [\beta\, (U-U^{-1})_{v_{0}v_{1}v_{2}}\,
(U-U^{-1})_{v_{0}v_{3}v_{4}} ]\;\; ,
\eea
for the $Z_{3}$ theory, and by
\be
B[v_{0},v_{1},v_{2},v_{3},v_{4}] = \exp[\beta(U-1)_{v_{0}v_{1}v_{2}}
(U-1)_{v_{0}v_{3}v_{4}}]\;\;,
\ee
for the $Z_{2}$ theory. Here, $U_{v_{0}v_{1}v_{2}} = U_{v_{0}v_{1}}
U_{v_{1}v_{2}}U_{v_{2}v_{0}}$ is the holonomy combination.
We refer to \cite{top1} for the precise normalization.

   To establish the triviality for closed $4$-manifolds,
consider the identity that was established in \cite{top2}; namely
that the Boltzmann weights of the $Z_{2}$ and $Z_{3}$ theories satisfy:
\begin{eqnarray}
1=& &W[0,1,2,3,4] \, W[0,1,2,4,5]\, W[0,1,2,5,3]\label{id} \\
& &W[1,0,3,4,5]\, W[2,1,3,4,5] \, W[0,2,3,4,5]\;\; ,\nonumber
\end{eqnarray}
at the special values of the coupling parameter. Written in this way, one
can recognize that the 4-simplices in this identity are actually
the boundary
of a 5-simplex $[0,1,2,3,4,5]$;
\begin{eqnarray}
\partial\, [0,1,2,3,4,5] =
& &[1,2,3,4,5] - [0,2,3,4,5] + [0,1,3,4,5] -\\
& &[0,1,2,4,5] + [0,1,2,3,5] - [0,1,2,3,4]\;\; .\nonumber
\end{eqnarray}
If $t$ is any 5-simplex, we can write this compactly as:
\begin{eqnarray}
W[\partial \, t] = 1\;\; .
\end{eqnarray}

    Let $K$ denote a simplicial complex which models a 4-manifold, possibly
with boundary. For our purposes, the 4-simplices $\{ s_{i} \}$ in K are
most important,
\begin{eqnarray}
K = \sum_{i} \; s_{i} \;\; .
\end{eqnarray}
Now consider the abstract simplicial complex called the cone over $K$
\cite{JM},
which is obtained by adding a new vertex $x$ to the simplicial complex $K$, and
linking it to all other vertices; we denote this simplicial complex by
$x \ast K$. Computing the boundary of that complex, one sees
\begin{eqnarray}
\partial \, (x \ast K) = K - x \ast \partial K \;\; .
\end{eqnarray}
Given that the Boltzmann weight of the left hand side is just 1, we have then
\begin{eqnarray}
W[K] = W[ x \ast \partial K]\;\; ,\label{cone}
\end{eqnarray}
where we mean, more precisely, that
\begin{eqnarray}
W[K] = \prod_{i} \; W[s_{i}] \;\; .
\end{eqnarray}
The simplicity of equation (\ref{cone}) is striking; the implication
is that the Boltzmann weight of any four dimensional simplicial complex
$K$ is identical to that of the cone over its boundary. Essentially, the
cone construction is giving a canonical presentation - or framing - of
the boundary of $K$. Moreover, if $\partial K$ has
several disjoint components
$M_{\alpha}$, then $K$ is a cobordism connecting them, and we immediately
have that
\begin{eqnarray}
Z[K] = \prod_{\alpha} \; Z[M_{\alpha}] \;\; .
\end{eqnarray}
This is one of the axioms for a topological field theory \cite{Atiyah}.

   Having established that we are dealing with a four dimensional gauge
theory which essentially reduces to something on the boundary, it is
natural to wonder about its interpretation as an intrinsically three
dimensional theory. As a gauge theory, we can gauge fix the links on any
maximal tree, and one such tree is given by the links which spew from
the vertex $x$. The value of the partition function is independent of how
we fix them, so we could always set those link variables to 1 say.
However one chooses to gauge fix these link variables, we can consider
the result to be a three dimensional lattice theory. If there is any
residual gauge invariance left, then it is not at all manifest, but
this is also reminiscent of the continuum Chern-Simons theories \cite{EWit}.

\section{Behavior Under Boundary Subdivision}

   In this section, we will undertake an analysis of the theory when a
bounding simplex is subdivided by an Alexander move of type 2 or 3. In
the later case,
a new vertex is added to the center of the 3-simplex and new links are
joined to the old vertices, yielding an assembly of four tetrahedrons.
We will find that the Boltzmann weight is not generally invariant under
this move, but we will be able to prove that this is so when we restrict
to field configurations which have trivial holonomy around all four faces
of the 3-simplex in question.

    Consider what happens to the Boltzmann weight of the theory when a
bounding 3-simplex $[0,1,2,3]$ is subdivided. Let $x$ be the cone vertex
discussed in the previous section, so that the partition function would
have the factor $W[0,1,2,3,x]$. Under subdivision where we add a new
vertex $c$ to the center of the tetrahedron, we would then consider a
new set of Boltzmann weights which are unchanged except that we would
replace the factor $W[0,1,2,3,x]$ by the quantity,
\begin{eqnarray}
W[c,1,2,3,x]\; W[0,c,2,3,x]\; W[0,1,c,3,x]\; W[0,1,2,c,x] \label{sdw}
\end{eqnarray}
in the partition function, and sum over the new link variables.
However, the main identity that we established
in \cite{top2}, namely (\ref{id}), says that this product of four weights is
equal to
\begin{eqnarray}
W[0,1,2,3,x]\; W^{-1}[0,1,2,3,c] \;\; .
\end{eqnarray}
Here, $W^{-1}$ denotes the inverse value which, for the Boltzmann weights
in our construction, is equivalent to simply an odd permutation of the
vertices; $W^{-1}[0,1,2,3,4] = W[0,1,2,4,3]$.
We see then that the subdivided Boltzmann weights represented by (\ref{sdw})
are precisely equivalent to having introduced an extra factor
$W^{-1}[0,1,2,3,c]$ into the original assembly of Boltzmann weights. It
is then crucial to understand how the theory behaves under insertions
of the kind:
\begin{eqnarray}
I[0,1,2,3] = \frac{1}{|G|^{4}}\; \sum_{U_{ci}} \;
W^{-1}[0,1,2,3,c]\;\; ,\label{insert}
\end{eqnarray}
where the sum is over the four link variables connected to $c$,
and $|G|$ is the order of the gauge group. At the
trivial points where (\ref{id}) holds, namely $s(2)=1$ or $s(3)=1$, this
quantity is manifestly 1. We are interested in investigating the
nontrivial roots of unity.

Let $U_{ijk} = U_{ij}U_{jk}U_{ki}$ denote the holonomy through the three
indicated vertices and $\delta(x)$ denote the usual delta function which is
1 for $x=0$ and zero otherwise.

{\bf Theorem 1:} The insertion $I[0,1,2,3]$ is equal to:
\begin{eqnarray}
\delta(U_{v_{0}v_{1}v_{2}}-1)\; \delta(U_{v_{0}v_{1}v_{3}}-1)\;
\delta(U_{v_{0}v_{2}v_{3}}-1) \;\delta(U_{v_{1}v_{2}v_{3}}-1)
\end{eqnarray}
at the points $s(2)=-1$, and $s(3)= \exp[\pm 2\pi i/3]$, in the $Z_{2}$
and $Z_{3}$ models respectively.

Thus, the insertion is 1 if all holonomies on the bounding 3-simplex
are trivial, and zero otherwise. This identity is most easily established
by computing both sides for all choices of boundary data and observing
that they are equal; we have done this with a computer program.

It is also interesting to consider Alexander
type 2 subdivision of a 2-simplex
which belongs to the bounding 3-manifold. Since the bounding space is
a manifold, a given 2-simplex, say $[0,1,2]$, will be shared by precisely
two 3-simplices; we denote their sum by $[0,1,2,3] - [0,1,2,4]$. The
requirement of the relative minus sign is dictated by the fact that we
have a 3-manifold without boundary. Under type 2 subdivision, we
add a new vertex $c$ to the center of the $[0,1,2]$ face, and link it
to the other vertices,
\begin{eqnarray}
[0,1,2] \rightarrow [c,1,2] + [0,c,2] + [0,1,c] \;\; .\label{t2}
\end{eqnarray}
Now, in the Boltzmann weights appropriate to the original complex, one
will find the product,
\begin{eqnarray}
W[0,1,2,3,x] \; W^{-1}[0,1,2,4,x]\;\; .
\end{eqnarray}
In the subdivided situation, each of these two factors will be replaced
by a product of three Boltzmann weights according to the structure
of (\ref{t2}). If we again use the identity (\ref{id}), one finds that
the subdivided situation is equivalent to the insertion of the following
factor in the original product of Boltzmann weights:
\begin{eqnarray}
I'[0,1,2,3,4] = \frac{1}{|G|^{5}}\; \sum_{U_{ci}}\;
W^{-1}[0,1,2,3,c] \; W[0,1,2,4,c] \;\; .
\end{eqnarray}

Let $U_{ijkl}= U_{ij}U_{jk}U_{kl}U_{li}$ denote the
holonomy through four vertices;
we then have the following result.

{\bf Theorem 2:} The quantity $I'[0,1,2,3,4]$ is equal to,
\begin{eqnarray}
\delta(U_{v_{0}v_{1}v_{2}}-1)\; \delta(U_{v_{0}v_{3}v_{1}v_{4}}-1)\;
\delta(U_{v_{1}v_{3}v_{2}v_{4}}-1)\;\delta(U_{v_{2}v_{3}v_{0}v_{4}}-1)\;\; ,
\end{eqnarray}
at the same points as in Theorem 1.

Notice that there is one 3-vertex holonomy around the 2-simplex $[0,1,2]$
which is the face common to the two 3-simplices that have been glued
together; the other 4-vertex holonomies are just products of the more
elementary holonomies. Since any 2-simplex on the boundary is common
to precisely two bounding 3-simplices, the restriction imposed by type
2 subdivision is actually equivalent to that from the type 3 move.

Again, the proof of Theorem 2 is most easily carried out with the aid
of a computer where one can simply compute both sides of the equation
for all values of boundary field configurations.

\section{Concluding Remarks}

As we have seen, achieving invariance with respect to
boundary subdivision
naturally leads to the insertion of operators which are delta
functions on trivial bounding holonomies. One might wonder if
these models
are somehow related to the discrete Chern-Simons type theories
of \cite{DW,Alt}, but we shall leave that issue open.

In \cite{top1}, the unrestricted partition function (where one includes
contributions from all bounding field configurations)
was evaluated for a simplicial
complex which consisted of a single $4$-simplex, thereby modelling the
$4$-disk. In addition, it was shown that the  results obtained
from such a computation were invariant under particular subdivisions
(single Alexander moves) of this particular complex. For the
case of the $Z_{2}$ and $Z_{3}$ models,
the partition function took the value $Z = 1/|G|^{3}$.
However, one should note that this value is
not generally invariant for arbitrary complexes which model the $4$-disk.
It suffices to consider a simple example.

Let us begin with a $4$-simplex $[0,1,2,3,4]$, and first perform an
Alexander  move of type $4$. This involves placing
a new vertex $5$ in the center of the simplex, and joining it to
all remaining vertices. The resulting complex takes the form:
\be
[5,1,2,3,4] + [0,5,2,3,4] + [0,1,5,3,4] + [0,1,2,5,4] + [0,1,2,3,5]\;\;.
\ee
A boundary subdivision is now induced by performing two Alexander
moves of type $3$; these are effected by introducing two additional
vertices, $6$ and $7$, in the center of the $3$-simplices
$[1,2,3,4]$ and $[0,1,3,4]$, respectively. The final complex is given by:
\bea
K&=& [5,6,2,3,4] + [5,1,6,3,4] + [5,1,2,6,4] + [5,1,2,3,6] \non\\
&+& [0,5,2,3,4] + [7,1,5,3,4] + [0,7,5,3,4] + [0,1,5,7,4] \non\\
&+& [0,1,5,3,7] + [0,1,2,5,4] + [0,1,2,3,5]             \;\;,
\eea
It is straightforward to compute the unrestricted partition function for $K$,
and the $Z_{2}$ model yields the result: $Z = 1/2^{5}$.

Given our analysis of boundary subdivisions, the central problem is to
construct an object related to the original partition function which
is fully subdivision invariant. This may involve, for example, restricting
the field configurations on the boundary. Alternatively, one might hope
that the subdivision moves eventually lead to a stable value for the
original partition function.

Another avenue which should be interesting to investigate concerns
modifications of the holonomy factors in our action. Since the $(U-1)$
combination essentially measures deviations from trivial holonomy,
one can simply consider replacing this by an angle, keeping the bow-tie
structure of the Boltzmann weight intact. This is in the spirit of the
Manton/Villain type actions. One would expect this simplification to
streamline the general analysis of these theories for all abelian groups.

On a different course, it is well known that $Z_{2}$ lattice gauge theory
of the ``$F^{2}$'' type is dual to the Ising model in 3 dimensions.
It would be interesting to investigate the duality properties of the
models we have been considering.

\end{document}